\documentclass[final,authoryear,5p,times,twocolumn]{elsarticle}

\usepackage{rotating}

\usepackage{graphics}
\usepackage{epstopdf}

\usepackage{graphicx}

\usepackage{amssymb}

\usepackage{natbib}

\begin{document}

\begin{frontmatter}



\title{Lorentz Symmetry Breaking studies with photons from Astrophysical Observations}



\author{J.~Bolmont and A.~Jacholkowska}
\address{LPNHE, Universit\'e Pierre et Marie Curie Paris 6, Universit\'e Denis Diderot Paris 7, CNRS/IN2P3, 4 Place Jussieu, F-75252, Paris Cedex 5, France}
\ead{bolmont@in2p3.fr, agnieszka.jacholkowska@cern.ch}

\begin{abstract}
Lorentz Invariance Violation (LIV) may be a good observational window on Quantum Gravity physics. Within last few years, all major gamma-ray experiments have published results from the search for LIV with variable astrophysical sources: gamma-ray bursts with detectors on-board satellites and Active Galactic Nuclei with ground-based experiments. In this paper, the recent time-of-flight studies with unpolarized photons published from the space and ground based observations are reviewed. Various methods used in the time delay searches are described, and their performance discussed. Since no significant time-lag value was found within experimental precision of the measurements, the present results consist of 95\% confidence level limits on the Quantum Gravity scale on the linear and quadratic terms in the standard photon dispersion relations.
\end{abstract}

\begin{keyword}
Lorentz Invariance \sep Quantum Gravity \sep Photon propagation \sep Gamma-ray bursts \sep Active galaxies
\end{keyword}

\end{frontmatter}

\parindent=0.5 cm

\tableofcontents

\section{Introduction}


Postulating various symmetries in Physics is largely justified by the quest for unified theories which rely on mathematical symmetries and on invariance of the fundamental laws. In the far past, the early Universe was dominated by a dynamical symmetry evoluting to the complex diversity of broken symmetries. In the history of the Universe the unique theory of Space and Time was characterized by Plank scales: $l_\mathrm{P} = (Gh/c^3)^{1/2} \approx 10^{-33}$ cm, \mbox{$t_\mathrm{P} \approx 10^{-43}$ s} and \mbox{$E_\mathrm{P} \approx 1.2\times10^{19}$ GeV}. Later, the emergence of new phenomena on hierarchical levels lead to our present world composed on one side of galaxies of stars and on small scales of atoms, molecules and life. At present there is no consensus on the correct approach to the unification of small and large-scale physics: the nature of the small-scale of the space-time could be derived from the idea of ``fuzziness'' (following to the Heisenberg uncertainty principle) or quantum foam discussed by \citet{wheeler},\citet{amelino98} or \citet{amelino02}, or sharply defined set of discrete points \citep{gambini}. This ``lumpiness'' or ``discreteness'' may affect photon propagation in the vacuum resulting in propagation anomalies. In consequence, the speed of light would differ with energy when photons travel through large distances. The searches for a physical theory uniformly valid in large (strong field gravity, High Energies, cosmology) and small scales (atoms, nuclei, elementary particles) need empirical tests across these scales. The quickly developing domain of studies with astronomical observations of distant sources may shed new light on the subject.

The purpose of this paper is purely experimental and no critical discussion about various theoretical approaches will be presented here, as it may be found in e.g. \citet[and references therein]{amelino09}. The formalism in use will be limited to the one adopted in the time-of-flight studies with unpolarized photons from distant astrophysical sources. These constraints are usually translated into Quantum Gravity Scale limits in the frame of considered models. In consequence, the naming of \textit{Quantum Gravity} will be used as a generic appellation of the quantum effects in the space-time structure.

The constraints on Lorentz Invariance Violation (LIV) derived here are much less competitive than those obtained in birefringence studies in the frame of the Standard Model Extensions scheme of \citet{colladay98,kostelecky08}. They should be considered as a complementary way to study the LIV effects and their consequences in the most model-independent way.

The Lorentz Symmetry has been successfully tested with experiments dealing with particle physics domain. However, there was no confirmation yet of its validity at large scales. Large theoretical interest in possible high energy violation of local Lorentz Invariance (LI) in the past decades was driven by possible hints for the Quantum Gravity (QG) as a local LI may not be an exact symmetry of the vacuum. As stated above, there are several reviews available on the subject \citep{amelino09,ellis09}.

The present review paper is organized as follows: the next section introduces the search for LIV and QG scale limits with astrophysical sources and briefly describes the formalism in use. Section \ref{sec:presdata} presents the results obtained with Pulsars, Gamma-ray Bursts and Active Galactic Nuclei. Finally, results from latest analyses are summarized and discussed in section~\ref{sec:conc}.

\section{Tests of Lorentz Invariance and search for Quantum Gravity with Astrophysics}

\subsection{Testing Lorentz invariance with astroparticles}

Several astrophysical messengers traveling from distant sources can be considered when testing LIV: photons, electrons, Ultra High Energy Cosmic Rays (UHECRs), neutrinos and gravitational waves. Each type of messenger pre\-sents its proper advantages and drawbacks. The propagation of photons can be simply described with dispersion relation formulas discussed below. On the other hand the detection of the high energy photons suffers from limitations in the energy lever arm due to the so called ``gamma horizon'' effect related to the interactions with infra-red background \citep{blanch}. The electrons emitted by pulsars synchrotron radiation allow very precise measurements of the time delays but their signals come from close-by sources \citep{jacobson06,liberati09,stecker09}. The UHECRs are currently used to investigate the possible modifications of the Greisen-Zatsepin-Kuzmin (GZK) threshold responsible for a cut-off in the UHECR spectrum at $\sim10^{18.5}$~eV \citep[see e.g.][]{sigl08,maccione09}. These studies provide already very stringent limits on LIV but are also very dependent on the particle physics modeling and on the anisotropy of their emission. The neutrinos from the astrophysical sources as well as the gravitational waves have not been yet observed. However, in case of a coincident detection, the results on LIV would be very stringent \citep{pradier09, elewyck09}. As recently remarked by several authors \citep[see e.g.][]{amelino09}, the LIV parameters may vary for different messengers as protons, electrons, gamma, neutrinos and gravitons, depending on a considered theoretical frame.

LIV studies with photons from cosmological sources were first proposed by \citet{amelino98}. As suggested in this article, the tiny effects due to quantized space-time can add up to measurable time delays for photons from distant sources. The hypothesis of energy dispersion could be best verified in sources that show fast flux variability, are at cosmological distances and are observed over a wide energy range between keV and TeV. This is the case of Gamma Ray Bursts (GRBs) and Very High Energy (VHE) flares of Active Galactic Nuclei (AGNs). Both types of sources are the preferred targets of these ``time-of-flight'' studies which provide the least model dependent tests of the Lorentz Symmetry breaking. The case of pulsed emission by Pulsars has also been considered, and provide valuable results as discussed below.

The measurement of the arrival time of photons in the detectors may only be possible if each trigger acquires a precise time-stamp provided by a spatial global positioning receiver. The precision in time below a micro-second, which is a key requirement of the described studies, allows a production of precise light curves for a given source on a few second time-scales. In particular, the Galileo network\,\footnote{http://www.satellite-navigation.eu} which is a global navigation satellite system (GNSS) currently being built by the European Union (EU) and the European Space Agency (ESA), will provide in future the needed accuracy of the detection time of individual photons. In addition, the UTC time provided by the GNSS makes possible the multi-wavelength analyses which combines results from various space experiments operating at lower energies such as SWIFT, INTEGRAL or the future satellite SVOM and higher energy satellites such as Fermi and AGILE, all integrated in the worldwide alert system (GCN).

The Figure of Merit of a given source for the time-lag measurement may be formulated as \citep{amelino98b}:
\begin{equation}
\label{eq:1}
\eta = \frac{L}{\mathrm{c}} \frac{E}{\mathrm{E}_\mathrm{QG}} \frac{1}{\Delta t},
\end{equation}
where $L$ is the distance of the source, $\Delta t$ is the measured time-lag and $\mathrm{E}_\mathrm{QG}$ is the Quantum Gravity scale.

\begin{figure*}
\begin{center}
\includegraphics[width=6in]{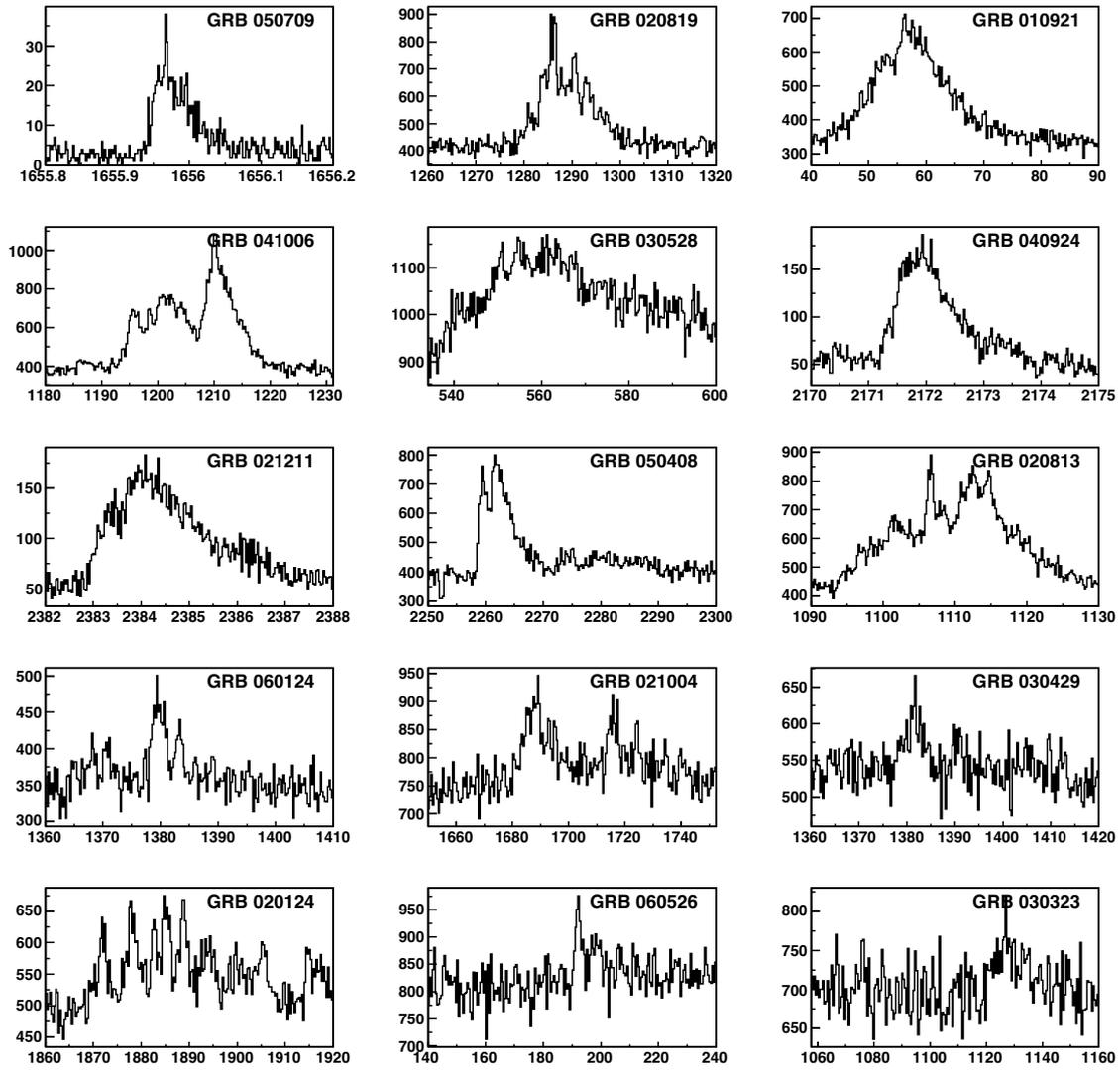}
\caption{Light curves of the 15 GRBs detected by HETE-2 in the energy range \mbox{6--400 keV} used in \citet{bolmont08}. The bursts are sorted by increasing $z$ from top left to bottom right. While $z$ increases, the signal-to-noise ratio decreases. $X$-axes are graduated in seconds.}
\label{fig:1}
\end{center}
\end{figure*}

\subsection{Measurement of the propagation effects with photons}

Following the formalism employed by \citet{jacob}, a general model in which the Lorentz Symmetry is broken at a very high energy denoted in the following by $\mathrm{E}_\mathrm{QG}$, is assumed. A natural value for the QG energy scale would be the Planck energy, however a large range of values should be considered.

For photons with much lower energies than $\mathrm{E}_\mathrm{QG}$, the first and second order corrections to the energy-momentum relation are generally taken into account, as shown with $n = \{1,2\}$ in the following equation:
\begin{equation}
\label{eq:2}
E^2 - p^2c^2 \simeq \pm p^2c^2 \left(\frac{pc}{\mathrm{E}_\mathrm{QG}}\right)^n.
\end{equation}

Considering sources at the cosmological distances, the analyses of time lags as a function of redshift requires a correction due to the expansion of the universe, which depends on the cosmological model. Following the analysis of the BATSE data and more recently of the HETE-2 and SWIFT GRB data \citep{bolmont08,ellis03,ellis06,ellis08a} and within a framework of the Standard Cosmological Model \citep{bahcall} with flat expanding universe and a cosmological constant, the difference in the arrival time of two photons with energy difference $\Delta E$ is:
\begin{equation}
\label{eq:3}
\Delta t = \mathrm{H}_0^{-1} \frac{\Delta E}{\mathrm{E}^l_\mathrm{QG}} \int_0^z \frac{(1 + z) dz}{h(z)}
\end{equation}
for a linear term and 
\begin{equation}
\label{eq:4}
\Delta t = \frac{3}{2} \mathrm{H}_0^{-1} \frac{\Delta E^2}{(\mathrm{E}^q_\mathrm{QG})^2} \int_0^z \frac{(1 + z)^2 dz}{h(z)},
\end{equation}
for a quadratic term, where $\Delta E^2$ is the quadratic energy difference. $h(z)$ is given by $h(z) = \sqrt{\Omega_\Lambda + \Omega_m (1+z)^3}$. $\Omega_m$, $\Omega_\Lambda$ and $\mathrm{H}_0$ are parameters of the Standard Cosmological Model ($\Omega_m = 0.3$, $\Omega_\Lambda = 0.7$ and $\mathrm{H}_0 = 70$ km\,s$^{-1}$\,Mpc$^{-1}$). The time-lag may decrease or increase with $\Delta E$ depending if the model lead to sub-luminal or super-luminal photons.

For completeness, it is necessary to mention that the formalism of Equations~\ref{eq:3} and~\ref{eq:4}, even if it has been commonly used by gamma-ray astrophysicists for more than a decade, is not the only one used in the community studying LIV. \citet{kostelecky08a} for example write these expressions as\,\footnote{Refer to Eq. 145 of \citet{kostelecky08a} for the exact expression of $Y_l$ and $Y_q$.}:
\begin{equation}
\label{eq:5}
\frac{\Delta t}{\Delta E} = Y_l \int_0^z \frac{(1 + z) dz}{\mathrm{H}_0\ h(z)}
\end{equation}
for a linear term and 
\begin{equation}
\label{eq:6}
\frac{\Delta t}{\Delta E^2} = Y_q \int_0^z \frac{(1 + z)^2 dz}{\mathrm{H}_0\ h(z)},
\end{equation}
for a quadratic term and set the limits on parameters $Y_l$ and $Y_q$. These parameters can easily be deduced from $\mathrm{E}^l_\mathrm{QG}$ and $\mathrm{E}^q_\mathrm{QG}$ by
\begin{equation}
\label{eq:7}
Y_l = \frac{1}{\mathrm{E}^l_\mathrm{QG}}
\end{equation}
and
\begin{equation}
\label{eq:8}
Y_q = \frac{3}{2} \frac{1}{(\mathrm{E}^q_\mathrm{QG})^2}.
\end{equation}
In the following, and as it is straightforward to switch from one formalism to the other, the one of Eqs.~\ref{eq:3} and~\ref{eq:4} will be used and the limits will be given on $\mathrm{E}^l_\mathrm{QG}$ and $\mathrm{E}^q_\mathrm{QG}$.

In order to probe the energy dependence of the velocity of light induced by LIV, analyses of time lags studied as a function of redshift with several sources are the best way to evaluate effects due to the photon propagation. In some cases, the parameter $K_l$ is defined as follows:
\begin{equation}
\label{eq:9}
K_l(z) = \int_0^z \frac{(1 + z) dz}{h(z)},
\end{equation}
to take cosmological effects into account, except for nearby sources for which the simple figure of merit (Eq.~\ref{eq:1}) is a good approximation. Till 2007 and the publications of \citet{jacob07,jacob} and \citet{bolmont08}, the formula in use in all presented results had a term $(1+z)$ missing under the integral, thus yielding underestimated values of limits on the Quantum Gravity scale. 

The energy-dependent time-lags may arise from propagation effects in the LIV scheme discussed in this paper. On the other hand, possible effects intrinsic to the astrophysical source emission could also produce energy-dependent time-lags. The analysis as a function of the redshift ensures in principle that the results are independent of source-induced time-lags. In the first order of the dispersion relations, the evolution of the time-lags as a function of $z$ can be written as:
\begin{equation}
\label{eq:10}
\langle\Delta t\rangle = a\,K_l(z) + b\,(1+z),
\end{equation}
where $a$ and $b$ parameters stand for extrinsic (Quantum Gravity) and intrinsic effects, respectively. When the redshift study is not possible, the intrinsic (or source) effects are treated in average or assumed to be negligible. This is the case for most results presented in Section~\ref{sec:presdata}.

The most expanding domain of LIV tests at the moment is related to the astrophysical experiments with photons. The space missions are equipped with excellent detectors for violent event detection such as pulsar and GRB emission. They included BATSE, HETE-2 and INTEGRAL in the past, SWIFT and Fermi at present. The ground based telescopes H.E.S.S., MAGIC, VERITAS and CANGAROO provide the highest $\Delta E$ with AGN flares but are limited in time variability.

\subsection{Methods}

There are various methods for the time-lag determination and subsequent derivation of the Quantum Gravity scale. In order to measure a tiny deviation from its standard value of the light velocity and reach $10^{19}$ GeV domain, the following experimental conditions should be fulfilled: in case of the GRBs where keV--MeV photons are detected, the precision on $\Delta t$ of the order of 10$^{-4}$ to 10$^{-3}$ s is needed. For GRBs and AGN flare photons in the GeV--TeV range, the required precision should reach a level of a second for an individual photon.

The procedures in use to determine time-lags rely on different statistical treatments of the data
and most of them follow advanced procedures: 
\begin{itemize}
 \item Cross Correlation Function (CCF) --- BATSE \citep{band}, H.E.S.S. \citep{aharonian08}, Fermi \citep{abdo09a, abdo09b};
 \item Energy Cost Function (ECF) --- MAGIC \citep{albert08}, Fermi \citep{abdo09a, abdo09b};
 \item Wavelet Transforms --- BATSE, HETE-2, SWIFT \citep{ellis03, ellis06, bolmont08}, H.E.S.S. \citep{aharonian08};
 \item Likelihood fit of QG model parameters --- INTEGRAL \citep{lamon08}, MAGIC \citep{martinez09}, H.E.S.S. \citep{abramowski10};
 \item Figure of merit formula (Eq.~\ref{eq:1}) --- Fermi \citep{abdo09a, abdo09b}.
\end{itemize}

For robust results, the use of at least two methods which probe different aspects of the light curves is recommended. In addition, careful error calibration studies by Monte Carlo simulations are mandatory for the extraction of the limits.

\begin{figure}
\begin{center}
\includegraphics[width=3.7in]{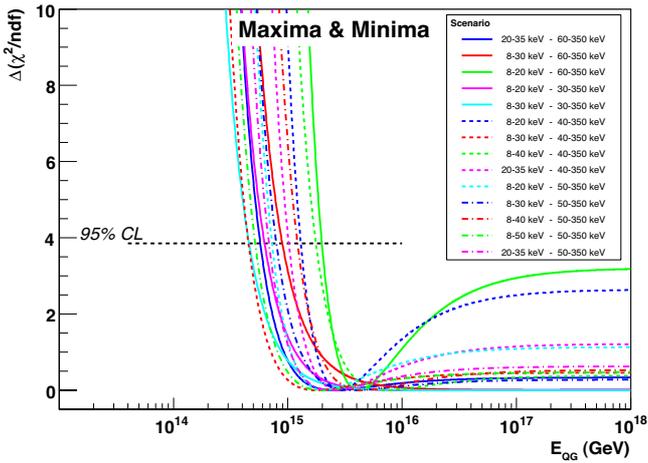}
\caption{Evolution of $\chi^2$ function of $\mathrm{E}^l_\mathrm{QG}$ for maxima and for minima found by CWT procedure in \citet{bolmont08}. Each curve corresponds to a different choice of the energy bands, and then to a value of $\Delta E$. All scenarios fulfill the condition \mbox{$\chi^2_\mathrm{min}(\mathrm{E}^l_\mathrm{QG})$/ndf $< 2$}.}
\label{fig:2}
\end{center}
\end{figure}

\section{Present data and results}\label{sec:presdata}

In this section the studies performed in view of search for Lorentz Symmetry breaking with Pulsars, GRBs and AGNs respectively are presented, following the best methods in use in this field. The analyses performed by the High Energy Transient Explorer (HETE-2), by Fermi GBM/LAT mission and by the H.E.S.S. and MAGIC experiments are good examples of current studies in presently running experiments. The quality and the significance of the results will be related to the acquired statistics and the type of method in use.

\subsection{Pulsars}

The pulsed emission from the Crab pulsar and nebula has been first studied by the EGRET experiment \citep{kaaret99} with respect to Quantum Gravity scale limit. Following to the alignment of the Crab pulsations (with 33.18~ms period) from radio to X-ray and gamma energies, a 95\% CL limit of \mbox{$1.8\times10^{15}$ GeV} was set on $\mathrm{E}_\mathrm{QG}$. This interesting result was achieved due to an excellent sub-millisecond precision of the EGRET clocks. A more stringent limit from the Crab pulsar is expected in future from the low energy threshold data acquired by the MAGIC collaboration.

\subsection{Gamma-Ray Bursts}
The GRBs are the most violent phenomena observed in the Universe, detected as sudden and unpredictable bursts of optical, hard X and $\gamma$-rays, lasting tenths of seconds and presenting variable and unclassified light curves. The GRBs are of cosmological origin and result from the death of a massive star or from the collapse of compact binary objects. Their very high variability of the order of milliseconds in the large energy range and distances going up to redshift values of 8, place them as excellent candidates for searches of the non-standard effects in the photon propagation.

An example of a typical analysis in the field in keV--MeV energy range has been performed by HETE-2 experiment. This mission was devoted to the study of GRBs using soft, medium X-rays and gamma-rays with instruments mounted on a compact spacecraft. The analysis of the 15 GRBs with measured redshifts collected by HETE-2 mission in years 2001-2006 followed the procedure described in details by \citet{bolmont08} and references therein. After the determination of the GRB interval time describing the start and the end of the burst, a cut above the background delimited the signal region to be studied in further analysis. The originality of the performed analysis was the choice of various energy ranges where the time-lags were computed with tagged photon data provided by the FREGATE sub-detector of the HETE-2 for each GRB. The light curves of the 15 GRBs are shown in Fig.~\ref{fig:1}. It appears clearly that the signal-to-noise ratio decreases for large redshifts. 

Following \citet{ellis03}, after a de-noising procedure of the light curves by a Discrete Wavelet Transform (DWT) \citep{donoho94}, a search for the rapid variations (spikes) in the light curves for all energy bands using a Continuous Wavelet Transform (CWT) \citep{mallat99} was performed. As a result, a list of minima and maxima candidates was followed by their association in pairs.

\begin{figure}
\begin{center}
\includegraphics[width=3.7in]{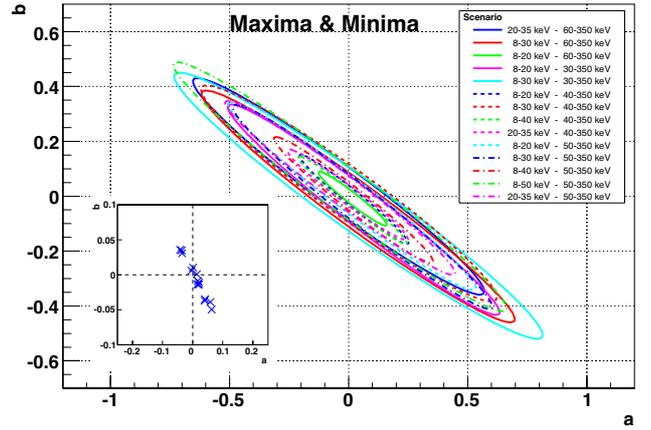}
\caption{95\% CL contours for $a$ and $b$ from the 2-parameter fits for the fourteen scenarios considered by \citet{bolmont08}. The box at the bottom left of the plots shows the position of contour centers.}
\label{fig:3}
\end{center}
\end{figure}

Fig.~\ref{fig:2} presents the evolution of $\chi^2(\mathrm{E}_\mathrm{QG})$ as computed with equation \ref{eq:10}, around its minimum $\chi^2_\mathrm{min}(\mathrm{E}_\mathrm{QG})$/ndf for maxima and minima together. The two-parameter linear fits, as discussed by \citet{bolmont08}, show a somewhat different behavior in case of the maxima and the minima. However, no significant preference of any value of $\mathrm{E}_\mathrm{QG}$ is observed for most of the scenarios when considering both types of extrema. 

Concerning possible source effects, it has been known for a long time, that the peaks of the emission in GRB light curves are shorter and arrive earlier at higher energies \citep{fenimore95,norris02}. These intrinsic lags, which have a sign opposite to the sign expected from LIV with subluminal photons, have a broad dispersion of durations, which complicates the detection of the Lorentz Symmetry violation effects. In the study with HETE-2 GRBs, a universality of the intrinsic source time-lags has been assumed. The 95\% CL contours for $a$ and $b$ from 2-dimensional fit are presented in Fig.~\ref{fig:3}, showing that both parameters representing the time-lags expected from the propagation and the source effects are strongly correlated. In this figure, the values of the offset parameter $b$ are compatible with zero for all energy scenarios. Both fit results suggest no variation above 2$\sigma$, so that the 95\% CL lower limits of the order of $4\times10^{15}$ GeV on the Quantum Gravity scale were derived.

\begin{figure}
\begin{center}
\includegraphics[width=3.2in]{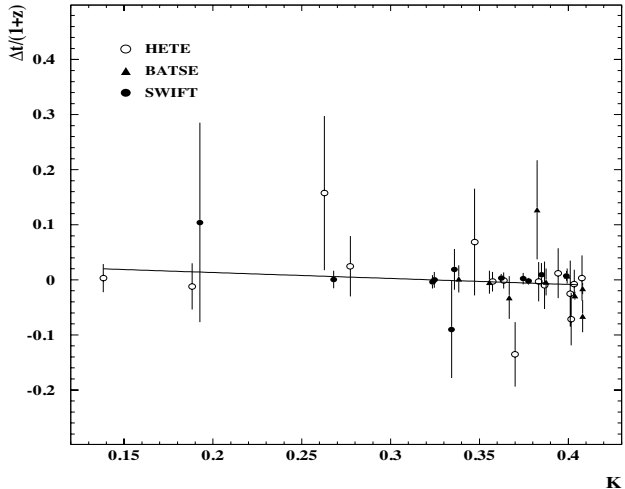}
\caption{Time-lags from high intensity GRBs as collected by BATSE, HETE-2 and SWIFT with rescaled errors, which lead to robust results, $\mathrm{E}^l_\mathrm{QG} > 2.1\times10^{16}$ GeV at 95\% CL \citep{ellis06,ellis08a}. $K$ is defined as in Eq.~\ref{eq:9}, but without the factor $(1+z)$ in the integral.}
\label{fig:4}
\end{center}
\end{figure}

In conclusion of the ``low energy'' studies, the time lags from GRB emission detected by HETE-2, BATSE, RHESSI, SWIFT and INTEGRAL space-borne experiments, as shown in Fig.~\ref{fig:4} for the measured time-lags, in the energy range of keV-MeV, provide robust lower limits on the Quantum Gravity scale of the order of $10^{16}$ GeV as analysed in \citep{ellis06,ellis08a}.

\begin{figure*}
\begin{center}
\includegraphics[width=6in]{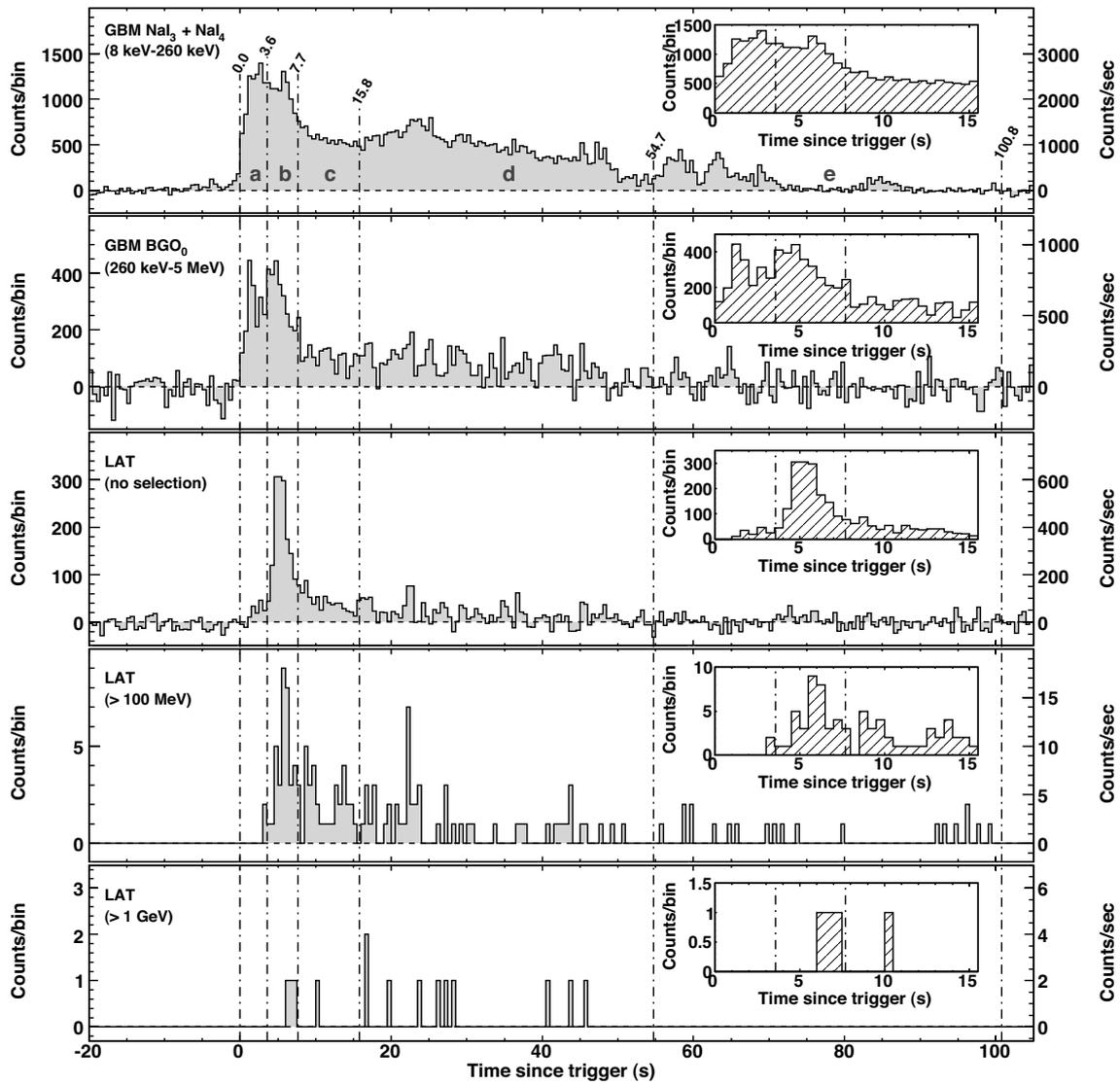}
\caption{Light curves in different energy bands for GRB 080916C as recorded by \textit{Fermi} instruments GBM (the two panels at the top) and LAT (the three panels at the bottom). Plot taken from \citet{abdo09a}.}
\label{fig:5}
\end{center}
\end{figure*}

\begin{figure*}
\begin{center}
\includegraphics[width=4.5in]{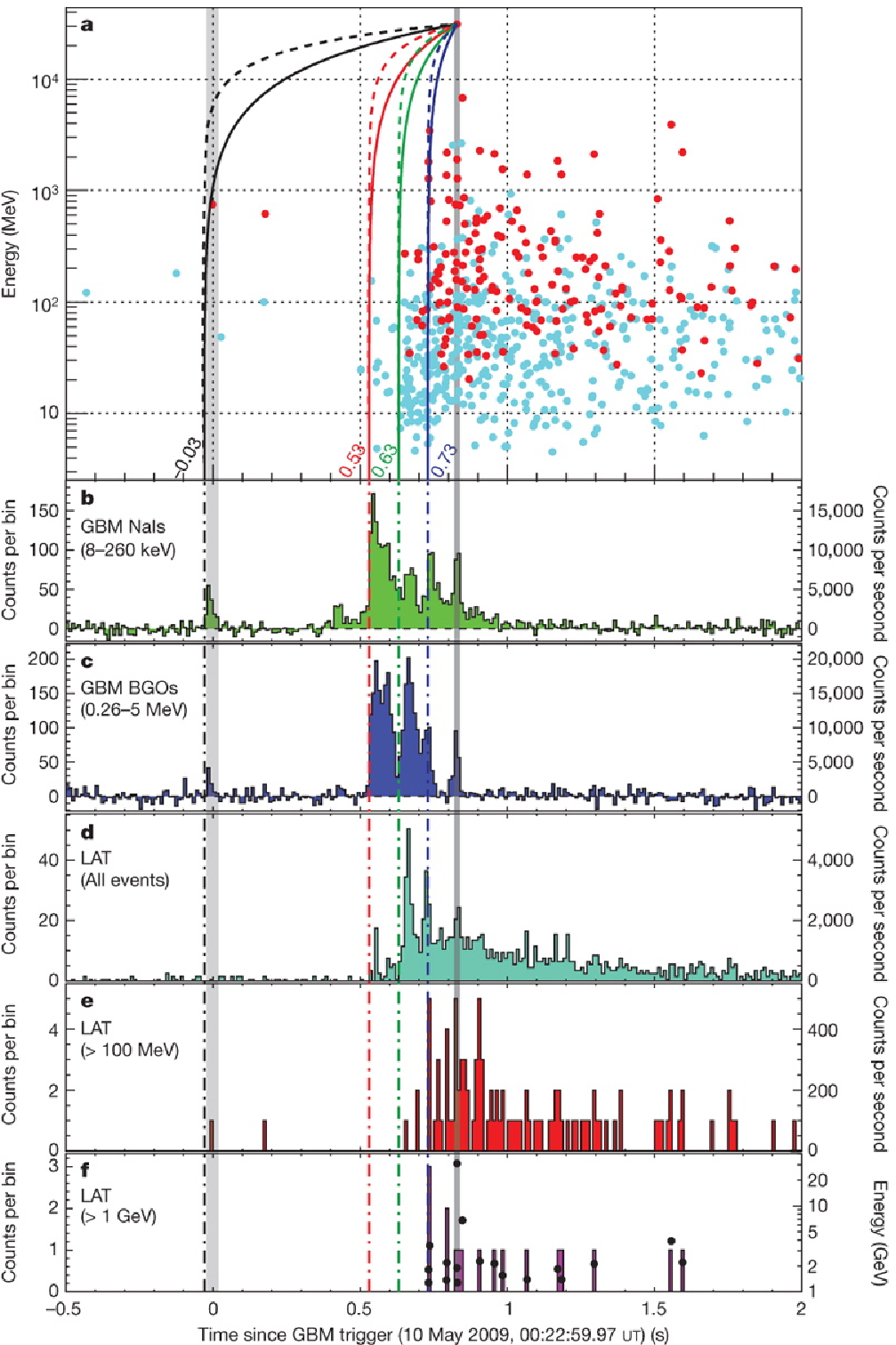}
\caption{Light curves in different energy bands for GRB 090510 \citep{abdo09b}. The top panel (a) shows the energy of each photon as a function of time. The vertical lines correspond to different hypotheses about when the highest energy photon (31~GeV) was emitted. These different hypotheses lead to limits on $\mathrm{E}^l_\mathrm{QG}$ varying from $1.19\,\mathrm{E}_\mathrm{P}$ to more than $100\,\mathrm{E}_\mathrm{P}$.}
\label{fig:6}
\end{center}
\end{figure*}

More recently, the Fermi collaboration \citep{abdo09a, abdo09b} has published two results on the Quantum Gravity scale lower limits reaching or exceeding the Planck scale from data taken with GBM and LAT sub-detectors, when detecting a powerful signal from GRB 080916C with a redshift value $\sim$4 and from a short burst (GRB~090510) at redshift of $\sim0.9$. The light curves as measured in energy bins are shown in Figs. \ref{fig:5} and~\ref{fig:6}. In both cases, substantial time delays of several seconds were measured between photons detected at keV (GBM) and GeV (LAT). As there is no strong evidence for a separate high energy component, the entire energy range was used to derive the limits on the LIV linear term of $\mathrm{E}^l_\mathrm{QG} > 1.5\times10^{18}$ GeV for the first GRB and $\mathrm{E}^l_\mathrm{QG} > 1.5\times10^{19}$ GeV at 95\% CL for the second one, attributing the measured lags to the source emission effects. These results constitute a breakthrough point in constraining the linear term models predicting Quantum Gravity scale below Planck scale. However, it should be noticed that both results rely on individual photons with energy $> 10$ GeV which determines the energy ranges. The limits were computed either by simply dividing $\Delta$(t) by the distance formula, or by CCF and Cost Function, with no statistical calibration procedure implied in the error estimation. To further progress in this domain of energy, future detection of several GRBs by Fermi mission with known redshift and measurable time-lags will be studied.

\subsection{Active Galaxies}

Blazars are variable AGNs, extragalactic sources producing $\gamma$-rays via the gravitational potential energy release of matter from an accretion disk surrounding a Super Massive Black Hole (SMBH). Beamed emission, large inferred luminosities, relativistic plasma jets pointing to the observer, and the flux variations by large factors (flares) on hour scale in time, make them excellent objects for variability studies from radio to VHE photons. In addition, the blazars are valuable transient candidates for searches of effects due to Lorentz Symmetry violation at Quantum Gravity scale. So far, more than 20 AGNs have been detected in VHE range with redshifts varying between 0.002 and 0.4. In spite of the fact that several AGNs have been already detected, only few provide enough luminosity for the analyses of their variability and photon propagation effects. Here, the case of the observation of an exceptional flare of the PKS 2155-304 ($z = 0.116$) by H.E.S.S. is described in more detail.
 
\begin{figure*}
\begin{center}
\includegraphics[width=4in]{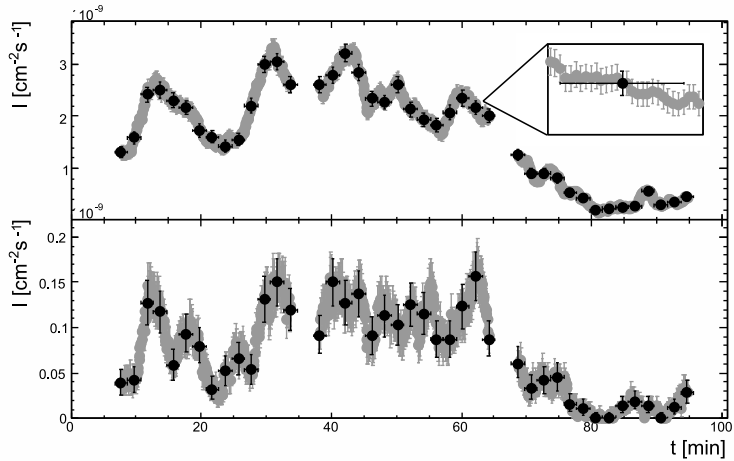}
\caption{Light curve of the PKS2155-304 flare during the night of 28th July 2006. The range of the energy is 200--800 GeV for the top panel and $> 800$ GeV for the bottom panel. The original data points (in black) are binned in 2-min time intervals. The zero time point is set to MJD 53944.02. Gray points show the over-sampled light curve, for which the 2-min bins are shifted in units of five seconds. Taken from \citet{aharonian08}.}
\label{fig:7}
\end{center}
\end{figure*}

In 2006, the H.E.S.S. experiment detected \citep{aharonian07} an exceptional flare of this source, with a high flux (10,000 photons recorded in 1.5 hours) and a high variability (rise and fall times of $\sim200$ s during the night of the 28th of July). The over-sampled light curve of the flare is shown in Fig.~\ref{fig:7} in two different energy bands.

To measure the time lags between photons in two different light curves, two independent analyses were carried out, using two different methods:
\begin{itemize}
 \item determining the position of the maximum of the Modified Cross Correlation Function (MCCF, \citet{li04}) which gives directly the value of the time lag. This method was applied to the over-sampled light curves of Fig.~\ref{fig:7} in the energy bands 0.2--0.8 TeV and $> 0.8$ TeV, which corresponds to a $\Delta E$ of 1.02 TeV.
 \item using a Continuous Wavelet Transform \citep{mallat99, bacry04} to locate with great precision the spike positions (extrema) in the light curves. An extremum of the low energy band was associated with an extremum in the high energy band to form a pair.
\end{itemize}

The fit of the MCCF curve with a Gaussian plus a first degree polynomial allowed for the determination of the time-lag between low and high energy photons of 20 s. In order to evaluate the uncertainties on this result, a detailed simulation of 10,000 light curves was performed in each energy band varying the flux within the error bars, leading to a calibrated error of 28 s. The obtained time-lag being found to be compatible with zero, a 95\% CL upper limit on the linear dispersion of 73 s/TeV was set.

With the CWT method, two pairs of extrema were obtained giving a mean time delay of 27 seconds. A similar method as the one used for the CCF was used to determine the errors and they were found to be in a range between 30 and 36 seconds. A 95\% confidence limit of 100 s/TeV was obtained for the linear correction to the dispersion relations. The MCCF method leads to a limit of the $\mathrm{E}^l_\mathrm{QG} > 7.2\times10^{17}$ GeV and the CWT to a confirmation of this value with a limit of the $\mathrm{E}^l_\mathrm{QG} > 5.2\times10^{17}$ GeV. The work in progress with Likelihood fit of the time-lag with individual photons provides a preliminary lower limit of $\mathrm{E}^l_\mathrm{QG} > 2\times10^{18}$ GeV \citep{abramowski10}, thus approaching considerably the limits found with Fermi GRBs.

\begin{figure}
\begin{center}
\includegraphics[width=3in]{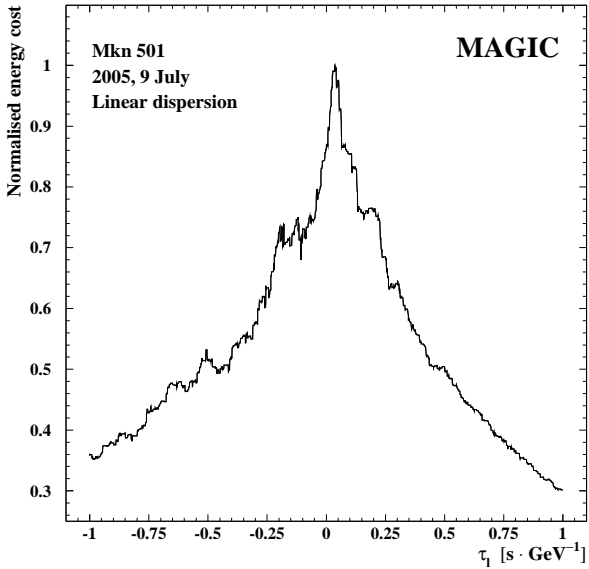}
\caption{Energy Cost Function (ECF) as a function of the time delay. The maximum value of the ECF provides the measured time-lag, here for the linear model. Figure taken from \citet{albert08}.}
\label{fig:8}
\end{center}
\end{figure}

The presented analysis with PKS 2155-304 flare can be compared with those performed by Whipple and MAGIC experiments with other AGN flares: Mkn~421 and Mkn~501. The Whipple collaboration has set a limit of \mbox{$4\times10^{16}$} GeV using a flare of Mkn~421 ($z = 0.031$) \citep{biller99}. More recently, the MAGIC collaboration obtained a limit of $3.2\times10^{17}$ GeV with a flare of Mkn~501 at $z = 0.034$ \citep{albert08}. Fig.~\ref{fig:8} presents the ECF plot as obtained by \citet{albert08} assuming a linear dependence on the Quantum Gravity scale. A similar value was derived from the same data with the likelihood fit method \citep{martinez09}. Although a nice minimum has been obtained in the likelihood function, a decisive conclusion could have not been established because of a limited statistics collected with Mkn~501 flare.

\begin{figure}
\begin{center}
\includegraphics[width=3in]{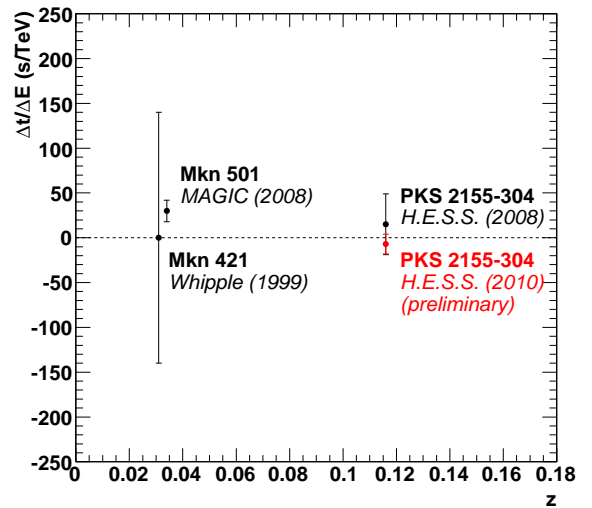}
\caption{The time-lag measured as a function of redshift for three active galaxies (Mkn 501, MKn 421 and PKS 2155-304). The results of the Whipple and MAGIC experiments are compared with the present H.E.S.S. results.}
\label{fig:9}
\end{center}
\end{figure}

The results obtained with H.E.S.S. are more constraining due to the fact that PKS~2155-304 is almost four times more distant than Mkn~421 and Mkn~501, and the acquired statistics were higher by a factor of ten. Fig.~\ref{fig:9} summarizes the results obtained so far with AGNs.

\section{Summary and discussion}\label{sec:conc}

Quantum Gravity phenomenology applied to studies with astrophysical sources has known a growing interest in the past decade, especially since it was argued that Lorentz Symmetry could be violated or at least distorted by the quantum nature of space-time, with measurable effects on photon propagation over large distances leading to a modification of the light group velocity.

The $\gamma$-ray dispersion studies in vacuum with GRBs and AGNs provide the cleanest way to look for Planck-scale modifications of the dispersion relations at the first order level. The results obtained with these sources for linear and quadratic corrections to the dispersion relations (translated into limits on parameters $\mathrm{E}^l_\mathrm{QG}$ and $\mathrm{E}^q_\mathrm{QG}$) are summarized in Table 1.

As discussed in this paper, the GRBs are good candidates for time of flight studies. The GRBs are easily detected by satellite experiments at very high redshifts up to $z \sim 8$ and up to a few hundreds of GeV in $\Delta E$. Population studies have already been carried out \citep{bolmont08,ellis06} and lead to limits of the order of $10^{16}$ GeV. The best and most spectacular limit so far is \mbox{$\mathrm{E}^l_\mathrm{QG} > 2\times10^{19}$ GeV} \citep{abdo09b}. The measurements of the AGN flares comes in complementarity to the GRB studies and yield at present lower limits on the linear term dispersion parameters only a factor of few below the Planck scale.

The existing synergy between studies done in Fermi experiment and ground based $\gamma$-ray telescopes opens new era in the presented studies. In principle, a consistency of the linear and quadratic term results derived with data from sources of different nature would provide convincing results on LIV phenomena. The coherence of the results in principle could be postulated. Alternatively, a persistence of different results with GRBs and AGNs may signify energy dependence of the effect. Combined analysis of GRB and AGN data may also lead to a better understanding of the source induced part of measured time-lags as it concerns analyses of sources at different redshifts. It has to be underlined here that a combined analysis needs adequate analysis methods and procedures for sources with different variability and different experimental setups.

Concerning the interpretation of the results, the most conservative one is to attribute the observed delays from MAGIC, H.E.S.S. and Fermi to the source emission effects having for its origin low energy gamma accelerated by electrons and high energy gamma by protons. The source delays are different for GRBs and AGNs (also different for short and long GRBs) and intrinsic fluctuations in the relative emission times for high and low components for sources in a given class or individual source have to be taken into account and have to be controlled as this aspect limits the sensitivity of these probes of the LIV effects. The only way at present to minimize these effects are population studies of GRBs and AGNs as a function of redshift which still need modelling of the source effects. The other important point concerns a potentiality of these studies to detect a propagation induced time-lag: this would be a much more difficult and challenging task due to the complexity of the studied emissions. In particular, there is no indication from the theory if the deviations are expected to be of the sub-luminal or super-luminal type, thus combining in different way with the source emission lags. Various proposals of studies are currently worked out as presented by \citet{amelino09}.

In conclusion, till now, no significant result on Lorentz Symmetry breaking has been obtained, analyzing large energy range for photons emitted by GRBs or Very High Energy data from flares of AGNs. The presented studies have shown a good sensitivity to the linear term in the photon dispersion relations, setting the lowest limit for the Quantum Gravity scale around $10^{19}$ GeV. To constrain at the same level the quadratic term would require the energies in the range of the Ultra High Energy Cosmic Rays (UHECRs) or extra-galactic neutrinos. In future, the redshift dependences will be explored to distinguish between intrinsic effects to the source or induced by Lorentz Symmetry Breaking. Further observations of both a high number of GRBs and of AGN flares will be necessary to give robust conclusions on possible propagation effects and possible deviations from Lorentz symmetry at high energies. Present and future experiments such as Fermi, CTA\footnote{http://www.cta-observatory.org/} or AGIS\footnote{http://www.agis-observatory.org/} will greatly improve our capabilities in this area.

\section*{Acknowledgments}

The authors would like to thank the organizers of the \textit{Scientific and Fundamental Aspects of the Galileo Programme} colloquium for their kind invitation. Special thanks go to the editor and referees for their helpful suggestions which greatly improved the quality of this review.

\clearpage

\begin{table*}
\caption{A selection of limits obtained with various instruments and methods for GRBs AGNs, and the Crab pulsar. Limits obtained for linear ($\mathrm{E}^l_\mathrm{QG}$) and quadratic ($\mathrm{E}^q_\mathrm{QG}$) corrections are given.\label{tab:1}}
\scriptsize
\begin{tabular}{llllll}
\hline
Source(s) & Experiment & Method & Results$^{a}$ & Reference & Note \\
\hline
GRB 021206   & RHESSI             & Fit + mean arrival time in a spike   & E$^{l}_\mathrm{QG} >$ 1.8$\times10^{17}$ GeV & \citet{boggs04}    & $^{b, c}$ \\
GRB 080916C  & Fermi GBM + LAT    & associating a 13 GeV photon with the  & E$^{l}_\mathrm{QG} >$ 1.3$\times10^{18}$ GeV & \citet{abdo09a}    &\\
             &                    & trigger time              &  E$^{q}_\mathrm{QG} >$ 0.8$\times10^{10}$ GeV   & &\\
GRB 090510   & Fermi GBM + LAT    & associating a 31 GeV photon with the  & E$^{l}_\mathrm{QG} >$ 1.5$\times10^{19}$ GeV & \citet{abdo09b}    & $^{d}$ \\
             &                    & start of any observed emission     & E$^{q}_\mathrm{QG} >$ 3.0$\times10^{10}$ GeV &   &\\
9 GRBs       & BATSE + OSSE       & wavelets                & E$^{l}_\mathrm{QG} >$ 0.7$\times10^{16}$ GeV & \citet{ellis03}    & $^{b}$ \\
             &                    &                     & E$^{q}_\mathrm{QG} >$ 2.9$\times10^{6}$ GeV &   &\\
15 GRBs      & HETE-2             & wavelets                & E$^{l}_\mathrm{QG} >$ 0.4$\times10^{16}$ GeV & \citet{bolmont08}   & $^{e}$ \\
17 GRBs      & INTEGRAL           & likelihood               & E$^{l}_\mathrm{QG} >$ 3.2$\times10^{11}$ GeV & \citet{lamon08}    & $^{f}$ \\
35 GRBs      & BATSE + HETE-2 + Swift    & wavelets                & E$^{l}_\mathrm{QG} >$ 1.4$\times10^{16}$ GeV & \citet{ellis06, ellis08a} & $^{g, h}$ \\
Mrk 421      & Whipple            & likelihood               & E$^{l}_\mathrm{QG} >$ 0.4$\times10^{17}$ GeV & \citet{biller99}    & $^{b, i}$ \\
Mrk 501      & MAGIC              & ECF                   & E$^{l}_\mathrm{QG} >$ 0.2$\times10^{18}$ GeV & \citet{albert08}    &\\
             &            &                     & E$^{q}_\mathrm{QG} >$ 2.6$\times10^{10}$ GeV &   &\\
             &            & likelihood               & E$^{l}_\mathrm{QG} >$ 0.3$\times10^{18}$ GeV & \citet{martinez09}   &\\
             &            &                     & E$^{q}_\mathrm{QG} >$ 5.7$\times10^{10}$ GeV &   &\\
PKS 2155-304 & H.E.S.S.        & MCCF                  & E$^{l}_\mathrm{QG} >$ 7.2$\times10^{17}$ GeV & \citet{aharonian08}  &\\
             &            &                     & E$^{q}_\mathrm{QG} >$ 0.1$\times10^{10}$ GeV &   &\\
             &            & wavelets                & E$^{l}_\mathrm{QG} >$ 5.2$\times10^{17}$ GeV &   &\\
             &            & likelihood               & E$^{l}_\mathrm{QG} >$ 2.1$\times10^{18}$ GeV  & \citet{abramowski10} &\\
             &            &                     & E$^{q}_\mathrm{QG} >$ 6.4$\times10^{10}$ GeV &   &\\
Crab pulsar  & EGRET         & shift of pulsation maxima in different & E$^{l}_\mathrm{QG} >$ 0.2$\times10^{16}$ GeV & \citet{kaaret99}  &\\
             &            & energy bands              &                       &   &\\
\hline
\end{tabular}

$^{a}$ Results can be expressed as by \citet{kostelecky08a} using the formulas $Y_l = 1/\mathrm{E}^l_\mathrm{QG}$ and $Y_q = 3/2 \cdot 1/(\mathrm{E}^q_\mathrm{QG})^2$.\\
$^{b}$ Limit obtained not taking into account the factor $(1+z)$ in the integral of Eq. \protect{\ref{eq:3}} and \protect{\ref{eq:4}}.\\
$^{c}$ The pseudo-redshift estimator \citep{pelan} was used. This estimator can be wrong by a factor of 2.\\
$^{d}$ Only the most conservative limit is given here.\\
$^{e}$ Photon tagged data was used.\\
$^{f}$ The pseudo-redshift estimator \citep{pelan} was used for 6 GRB out of 11.\\
$^{g}$ For HETE-2, fixed energy bands were used.\\
$^{h}$ The limits of \citet{ellis06} were updated in \citet{ellis08a} taking into account the factor $(1+z)$ in the integral of Eq. \protect{\ref{eq:3}} and \protect{\ref{eq:4}}. Only the limit obtained for a linear correction is given.\\
$^{i}$ A likelihood procedure was used, but not on an event-by-event basis.

\end{table*}



\begin{thebibliography}{1}


\bibitem[Abdo et al.(2009a)]{abdo09a} Abdo, A.A.,Ackermann, M., Arimoto, M. et al. (Fermi Collaboration), Fermi Observations of High-Energy Gamma-Ray Emission from GRB 080916C. Science Express, 19.02.2009a.
\bibitem[Abdo et al.(2009b)]{abdo09b} Abdo, A.A.,Ackermann, M., Ajello, M. et al. (Fermi Collaboration). A limit on the variation of the speed of light arising from quantum gravity effects. Nature 462, 331--334, 2009b.
\bibitem[Abramowski et al.(2011)]{abramowski10} Abramowski, A., Acero, F., Aharonian, F. et al. (H.E.S.S. Collaboration). Search for Lorentz Invariance breaking with a likelihood fit of the PKS2155-304 Flare Data Taken on MJD~53944. Accepted for publication in Astropart. Phys. Available from $<$arXiv:1101.3650$>$, 2011.
\bibitem[Aharonian et al.(2008)]{aharonian08} Aharonian, F., Akhperjanian, A.G., Barres de Almeida, U. et al. (H.E.S.S. Collaboration). Limits on an Energy Dependence of the Speed of light from a flare of the Active Galaxy PKS 2155-304. Phys. Rev. Lett. 101, 170402, 2008.
\bibitem[Aharonian et al.(2007)]{aharonian07} Aharonian, F., Akhperjanian, A.G., Bazer-Bachi, A.R. et al. (H.E.S.S. Collaboration). An Exceptional VHE Gamma-Ray Flare of PKS 2155-304. Astrophys. Journal Lett. 664, L71--L74, 2007.
\bibitem[Albert et al.(2008)]{albert08} Albert, J., Aliu, E., Anderhub, L.A. et al., (MAGIC Collaboration) and Ellis, J., Mavromatos, N.E., Nanopoulos, D.V. et al. Probing quantum gravity using photons from a flare of the active galactic nucleus Markarian 501 observed by the MAGIC telescope. Phys. Lett. B 668, 253--257, 2008.
\bibitem[Amelino-Camelia(2002)]{amelino02} Amelino-Camelia, G. Quantum gravity phenomenology: Status and prospects. Mod. Phys. Lett. A17, 899--922, 2002.
\bibitem[Amelino-Camelia et al.(1998a)]{amelino98} Amelino-Camelia, G., Ellis, J., Mavromatos, N.E. et al. Tests of quantum gravity from observations of $\gamma$-ray bursts. Nature 395, 525--525, 1998.
\bibitem[Amelino-Camelia et al.(1998b)]{amelino98b} Amelino-Camelia, G., Ellis, J., Mavromatos, N.E. et al. Sensitivity of Astrophysical Observations to Gravity-Induced Wave Dispersion in Vacuo. Available from $<$arXiv:astro-ph/9810483$>$, 1998.
\bibitem[Amelino-Camelia and Smolin(2009)]{amelino09} Amelino-Camelia, G. and Smolin, L. Prospects for constraining quantum gravity dispersion with near term observations. Phys. Rev. D 80, 084017, 2009.
\bibitem[Bacry(2004)]{bacry04} Bacry, H., LastWave version 2.0.3, 2004.
\bibitem[Bahcall et al.(1999)]{bahcall} Bahcall, N.A., Ostriker, J.P., Perlmutter, S. et al. The Cosmic Triangle: Revealing the State of the Universe. Science 284, 1481--1488, 1999.
\bibitem[Band(1997)]{band} Band, J.P., Gamma-Ray Burst Spectral Evolution Through Cross-correlations of Discriminator Light Curves. Astrophys. Journal 486, 928--937,1997.
\bibitem[Biller et al.(1999)]{biller99} Biller, S.D., Breslin, A.C., Buckley, J. et al. (Whipple Collaboration). Limits to Quantum Gravity Effects from Observations of TeV Flares in Active Galaxies. Phys. Rev. Lett. 83, 2108--2111, 1999.
\bibitem[Blanch and Martinez(2005)]{blanch} Blanch, O. and Martinez M. Exploring the gamma-ray horizon with the next generation of gamma-ray telescopes. Part 2: Extracting cosmological parameters from the observation of gamma-ray sources. Astropart. Phys. 23, 598--607, 2005.
\bibitem[Boggs et al.(2004)]{boggs04} Boggs, S.E., Wunderer, C.B., Hurley, K. et al. Testing Lorentz Invariance with GRB 021206. Astrophys. Journal 611, L77-L80, 2004.
\bibitem[Bolmont et al.(2008)]{bolmont08} Bolmont, J., Jacholkowska, A., Atteia, J.-L. et al. Study of time lags in HETE-2 Gamma-Ray Bursts with redshift : search for astrophysical effects and Quantum Gravity signature. Astrophys. Journal 676, 532--544, 2008.
\bibitem[Colladay and Kosteleck\'y(1998)]{colladay98} Colladay, D. and Kosteleck\'y, V.A. Lorentz-violating extension of the standard model. Phys. Rev. D  58, 116002, 1998.
\bibitem[Donoho and Johnstone(1994)]{donoho94} Donoho, D.L. and Johnstone, J.M. Ideal spatial adaptation by wavelet shrinkage. Biometrika, 81, 425--455, 1994.
\bibitem[Ellis et al.(2003)]{ellis03}   Ellis, J., Mavromatos, N.E., Nanopoulos, D.V. et al. Quantum-Gravity Analysis of Gamma-Ray Bursts using Wavelets. Astronomy and Astrophysics, 402, 409--424, 2003.
\bibitem[Ellis et al.(2006)]{ellis06}   Ellis, J., Mavromatos, N.E., Nanopoulos, D.V. et al. Robust limits on Lorentz violation from gamma-ray bursts. Astroparticle Phys. 25, 402--411, 2006.
\bibitem[Ellis et al.(2008)]{ellis08a} Ellis, J., Mavromatos, N.E., Nanopoulos, D.V. et al. Corrigendum to ``Robust limits on Lorentz violation from gamma-ray bursts''. Astropart. Phys., 29, 158--159, 2008.
\bibitem[Ellis et al.(2009)]{ellis09} Ellis, J., Mavromatos, N.E. and Nanopoulos, D.V. Probing a Possible Vacuum Refractive Index with Gamma-Ray Telescopes. Phys. Lett. B, 674, 83--86, 2009.
\bibitem[Fenimore et al.(1995)]{fenimore95} Fenimore, E. E., in't Zand, J.J.M., Norris, J.P. et al. Gamma-ray burst peak duration as a function of energy. Astrophys. Journal 448, L101--L104, 1995.
\bibitem[Galaverni and Sigl(2008)]{sigl08} Galaverni, M. and Sigl, G. Lorentz Violation for Photons and Ultra-High Energy Cosmic Rays. Phys. Rev. Lett., 100, 021102, 2008.
\bibitem[Gambini and Pullin(1999)]{gambini} Gambini, R. and Pullin, J. Non standard optics from quantum space-time. Phys. Rev. D 59, 124021, 1999.
\bibitem[Jacob and Piran(2007)]{jacob07} Jacob, U. and Piran, T. Neutrinos from gamma-ray bursts as a tool to explore quantum-gravity-induced Lorentz violation. Nature Phys. 3, 87--90,2007.
\bibitem[Jacob and Piran(2008)]{jacob} Jacob, U. and Piran, T. Lorentz-violation-induced arrival delays of cosmological particles. Journal of Cosmology and Astro-Particle Physics 01, 031, 2008.
\bibitem[Jacobson et al.(2006)]{jacobson06}Jacobson, T., Liberatti, S. and Mattingly D., Lorentz violation at high energy: concepts, phenomena and astrophysical constraints. Annals Phys., 321, 150--196, 2006.
\bibitem[Kaaret(1999)]{kaaret99} Kaaret, P., Pulsar Radiation and Quantum Gravity. Astron. Astrophys. 345, L32--L34, 1999.
\bibitem[Kosteleck\'y and Mewes(2008)]{kostelecky08a} Kosteleck\'y, A. and Mewes, N. Electrodynamics with Lorentz-violating operators of arbitrary dimension. Phys. Rev. D 80, 015020, 2009.
\bibitem[Kosteleck\'y and Russell(2008)]{kostelecky08} Kosteleck\'y, A. and Russell, N. Data Tables for Lorentz and CPT Violation. Available from $<$arXiv:0801.0287$>$, 2008.
\bibitem[Lamon et al.(2008)]{lamon08} Lamon, P., Produit, N. and Steiner, F. (INTEGRAL Collaboration). Study of Lorentz violation in INTEGRAL gamma-ray bursts. Gen. Rel. Grav. 40, 1731--1743, 2008.
\bibitem[Li et al.(2004)]{li04} Li, T.-P., Qu, J.-L., Feng, H. et al. Timescale Analysis of Spectral Lags. Chinese Journal of Astronomy and Astrophys. 4, 583--598, 2004.
\bibitem[Liberati and Maccione(2009)]{liberati09} Liberati, S., Maccione, L. Lorentz Violation: Motivation and New Constraints. Ann. Rev. Nucl. Part. Sci. 59, 245--267, 2009.
\bibitem[Maccione et al.(2009)]{maccione09} Maccione, L., Taylor, A.~M., Mattingly, D.~M. et al. Planck-scale Lorentz violation constrained by Ultra-High-Energy Cosmic Rays. Journal of Cosmology and Astro-Particle Physics 04, 022, 2009.
\bibitem[Mallat(1999)]{mallat99} Mallat, S. A Wavelet Tour of Signal Processing. Academic Press, London, 1999.
\bibitem[Martinez and Errando(2009)]{martinez09} Martinez, M. and Errando, M. A new method to study energy-dependent arrival delays on photons from astrophysical sources. Astropart. Phys. 31, 226--232, 2009.
\bibitem[Norris(2002)]{norris02} Norris, J. P. Implications of Lag-Luminosity Relationship for Unified GRB Paradigms. Astrophys. Journal 579, 386--403, 2002.
\bibitem[P\'elangeon et al.(2006)]{pelan} P\'elangeon, A., Atteia, J.-L., Lamb, D.Q. et al. An improved redshift indicator for Gamma-Ray Bursts, based on the prompt emission. In {\em Gamma-Ray Bursts in the Swift Era}, AIP Conf. Proc. vol. 836, 149--152, 2006.
\bibitem[Pradier(2009)]{pradier09} Pradier, T. Coincidences between gravitational wave interferometers and high energy neutrino telescopes. Nuclear Instruments and Methods in Physics Research A 602, 268--274, 2009.
\bibitem[Stecker and Scully(2009)]{stecker09} Stecker, F.W., Scully, S.T. Searching for new physics with ultrahigh energy cosmic rays. New J. Phys. 11, 085003, 2009.
\bibitem[van Elewyck et al.(2009)]{elewyck09} van Elewyck, V., Ando, S., Aso, Y. Joint Searches Between Gravitational-Wave Interferometers and High-Energy Neutrino Telescopes: Science Reach and Analysis Strategies. International Journal of Modern Physics D 18, 1655--1659, 2009.
\bibitem[Wheeler(1982)]{wheeler} Wheeler, J.A. The computer and the universe. Int. J. Theor. Phys. 21, 557--572, 1982.





\end{thebibliography}
\end{document}